\documentclass[aps,prb,preprint,superscriptaddress]{revtex4-1}
\usepackage[applemac]{inputenc}	
\usepackage[T1]{fontenc}
\usepackage[cyr]{aeguill}
\usepackage[pdftex]{graphicx}
\usepackage{wrapfig}
\usepackage[english,french]{babel}
\usepackage{amsmath}
\usepackage{amssymb,amsfonts,textcomp}
\usepackage{color}
\usepackage{array}
\usepackage{hhline}
\usepackage{hyperref}
\usepackage{fancyhdr}
\usepackage[pdftex]{graphicx}
\definecolor{violet}{rgb}{0.5,0,0.5}
\definecolor{violet2}{rgb}{0.61,0,0.87}
\definecolor{vert}{rgb}{0,0.65,0}
\definecolor{vert2}{rgb}{0,0.6,0}
\definecolor{marron}{rgb}{0.79,0.52,0}
\definecolor{orange}{rgb}{0.87,0.47,0}
\definecolor{lightblue}{rgb}{0,0.8,0.8}
\definecolor{jaune}{rgb}{0.948,0.91,0.17}

\begin{document}
\thispagestyle{empty}
\title{Strain superlattices and macroscale suspension of Graphene induced by corrugated substrates}
\author{Antoine Reserbat-Plantey}\altaffiliation{Present address : ICFO-Institut de Ciencies Fotoniques, Mediterranean Technology Park, 08860 Castelldefels (Barcelona), Spain.}
\affiliation{Univ. Grenoble Alpes, CNRS, I. Neel, F-38000 Grenoble, France}
\author{Dipankar Kalita}\affiliation{Univ. Grenoble Alpes, CNRS, I. Neel, F-38000 Grenoble, France}
\author{Laurence Ferlazzo}\affiliation{Laboratoire de photonique et de Nanostructures, CNRS, Marcoussis, France}
\author{Sandrine Autier-Laurent}\affiliation{Laboratoire de Physique des Solides, Université Paris-Sud-CNRS, Orsay, France}
\author{Katsuyoshi Komatsu}\affiliation{Laboratoire de Physique des Solides, Université Paris-Sud-CNRS, Orsay, France}
\author{Chuan Li}\affiliation{Laboratoire de Physique des Solides, Université Paris-Sud-CNRS, Orsay, France}
\author{Raphaël Weil}\affiliation{Laboratoire de Physique des Solides, Université Paris-Sud-CNRS, Orsay, France}
\author{Zheng Han}\affiliation{Univ. Grenoble Alpes, CNRS, I. Neel, F-38000 Grenoble, France}
\author{Sandrine Autier-Laurent}\affiliation{Laboratoire de Physique des Solides, Université Paris-Sud-CNRS, Orsay, France}
\author{Arnaud Ralko}\affiliation{Univ. Grenoble Alpes, CNRS, I. Neel, F-38000 Grenoble, France}
\author{Laetitia Marty}\affiliation{Univ. Grenoble Alpes, CNRS, I. Neel, F-38000 Grenoble, France}
\author{Sophie Guéron}\affiliation{Laboratoire de Physique des Solides, Université Paris-Sud-CNRS, Orsay, France}
\author{Nedjma Bendiab}\affiliation{Univ. Grenoble Alpes, CNRS, I. Neel, F-38000 Grenoble, France}
\author{Hélène Bouchiat}\affiliation{Laboratoire de Physique des Solides, Université Paris-Sud-CNRS, Orsay, France}
\author{Vincent Bouchiat}\affiliation{Univ. Grenoble Alpes, CNRS, I. Neel, F-38000 Grenoble, France}


\begin{abstract}
We investigate the organized formation of strain, ripples and suspended features in macroscopic CVD-prepared graphene sheets transferred onto a corrugated substrate made of an ordered arrays of silica pillars of variable geometries. 
Depending on the aspect ratio and sharpness of the corrugated array, graphene can conformally coat the surface, partially collapse, or lay, fakir-like, fully suspended between pillars over tens of micrometers.  
Upon increase of pillar density, ripples in collapsed films display a transition from random oriented pleats emerging from pillars to ripples linking nearest neighboring pillars organized in domains of given orientation.  Spatially-resolved Raman spectroscopy, atomic force microscopy and electronic microscopy reveal uniaxial strain domains in the transferred graphene, which are induced and controlled by the geometry.  
We propose a simple theoretical model to explain the transition between suspended and collapsed graphene. 
For the arrays with high aspect ratio pillars, graphene membranes stays suspended over macroscopic distances with minimal interaction with pillars tip apex. It offers a platform to tailor stress in graphene layers and open perspectives for electron transport and nanomechanical applications.  
\bigskip
\end{abstract}
\date{\today}
\maketitle
\clearpage
\textbf{}


\section{Introduction}
Graphene, the two-dimensional honeycomb carbon lattice, has unique mechanical properties such as strong in-plane rigidity together with a huge elasticity range as it can withstand up to 25\% elastic deformation \cite{Lee2008}. 
It is today the only atomically-thin material that routinely provides stable and self-supported membranes, allowing a wide range of applications ranging from nanoelectonic and optomechanical devices to biology: among notable recents results involving graphene membrane as its critical component, one can cite  high electronic mobility devices showing fractional quantum hall effect \cite{Du2009},  nano-electromechanical systems\cite{Zande2010}, leak-proof membrane\cite{Bunch2008}, offering promising materials  for water filtration\cite{Nair2012} and DNA sequencing\cite{Garaj2010}.

The development of graphene growth over centimeter scale area and the improvement of transfer techniques make it all the more important to control the shape and geometry of graphene once transferred onto the destination substrate. 
Indeed, the possibility of growing continuous monolayer graphene\cite{Li2009-2,Kim2009,Han2012,Gao2012} onto sacrificial catalytic layers has enabled manipulation of large areas of graphene and makes possible its transfer onto surfaces of arbitrary shape and composition.  
Once transferred on a flat surface, or further suspended\cite{Bao2009,Meyer2007,Fasolino2007}, graphene membranes always display unwanted ripples that affect its electrical\cite{Ni2012}, thermal\cite{Chen2009b} and mechanical\cite{Bao2012} properties. 
Wrinkles (reminiscent as the one occurring in hanging draperies) that  develop in doubly-clamped graphene membranes under uniaxial stress\cite{Vandeparre2011} induce additional damping in electromechanical systems \cite{Bao2009}, whereas ripples in graphene-based transistors are known alter the electrical conductivity \cite{Ni2012}.
Nevertheless these mechanical-induced defects can be sometimes desirable as a means to engineer a controlled level of stress either to generate an electrical gap\cite{Pereira2009,Guinea2009,Covaci2012} or to induce strong pseudo-magnetic fields\cite{Levy2010,Neek-Amal2012a}.

Before reaching such a stage of control, it appears necessary to understand the interaction process between a polycrystalline graphene membrane and the destination substrate onto which it is wet-transferred.
In this paper, we investigate the formation process of strain ripples and suspended features in graphene layers obtained by chemical vapor deposition on copper and transferred onto a corrugated substrate formed by an array of SiO$_2$ nano-pillars.
We show how to engineer the formation of graphene ripples using an ordered corrugated substrate which defines self-organized strain domains forming sets of parallel ripples linking the pillars.  By tuning the aspect ratio of the pillars from the array and its apex sharpness, we show that different membrane shape regimes can be reproducibly found. 
We explore both limits of low density arrays where graphene exhibits ripples domains and of very dense arrays for in which graphene does not ripple, but on the contrary stays fully suspended, fakir-like, over a dense array of nano-pillars (cf. \ref{fig1}).


\section{Sample preparation}
\begin{figure}[htbp]
\begin{center}
		\includegraphics[width=0.7\textwidth]{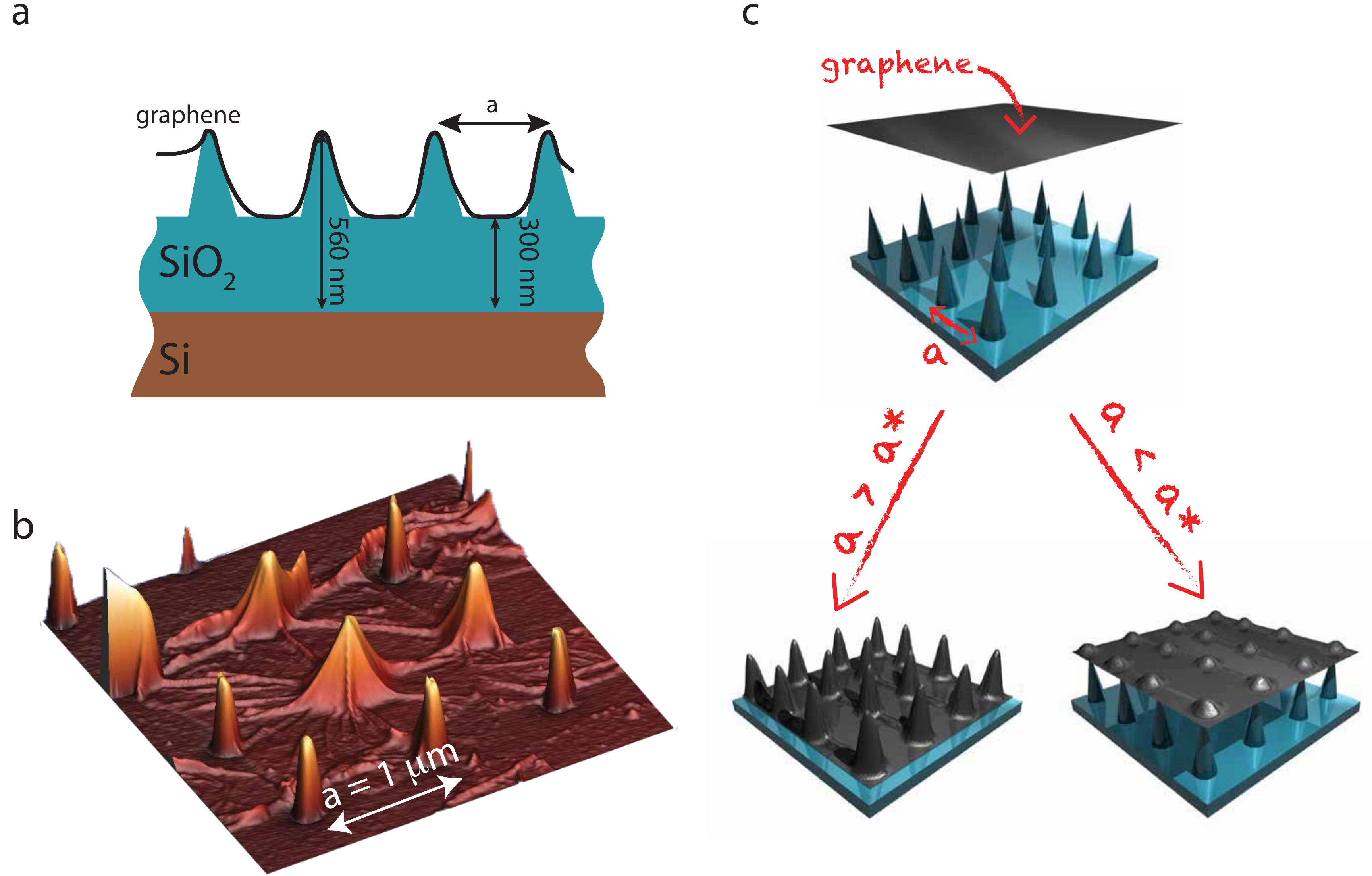}
	     \caption{\textbf{Transferred Graphene on nano-pillars.} \textbf{a}: Schematic view of graphene membrane deposited onto SiO$_2$ nano-pillar array. \textbf{b}: Atomic force micrograph of graphene deposited on SiO$_2$ nano-pillars.\textbf{c}: cartoon of graphene (in black) transferred onto nano-pillars array (in blue). For dense array ($a<a^*$), we observe fully suspended graphene over large areas. At low array density ($a>a^*$), graphene fits the substrate and forms highly symmetric ripples.}\label{fig1}
\end{center}
\end{figure}

The graphene sheets are obtained by chemical vapor deposition (CVD) growth on a sacrificial copper foil as described in our previous work\cite{Han2012}. 
This growth method produces homogeneous mono-layer graphene sample at the centimeter scale, with no second layer and polycrystalline film with perfectly stitched crystal grain of typical size 20 microns. 
The detailed fabrication process of nano-pillar array is presented in the supplementary informations. 
A PMMA coating film is used as a flexible and supporting layer to carry the graphene and deposit it onto the structured substrate, after acid etching of the copper catalyst (cf. Supp. Info.). 
The transfer is then realized by slowly picking up from below the PMMA/Graphene layer with the clean nano-pillar substrate, followed by a natural drying in air for one hour. 
Because residual liquid may be trapped under graphene, and to increase the chance of sticking onto the substrate, the whole sample was soft-baked before removing PMMA using acetone.
The final structure consists of a monolayer graphene sheet on a SiO$_2$ nano-pillar array of variable lattice parameter $a$.
Each nano-pillar is about 260 nm high, and the distance $a$ between two pillars varies from 250 nm to 3 $\mu$m (cf. \ref{fig1}).
Our method differs significantly from the one reported by another study\cite{Tomori2011} which involved suspension of graphene over pillar arrays after transfer, by in-situ releasing the polymer membrane using etching through the graphene layer. 
Similar systems of graphene on pillars have been studied previously\cite{Tomori2011,Lee2010,Lee2011} mostly using pillar arrays with flat, mesa-like ends. 
In the work of Tomori \textit{et al}\cite{Tomori2011}, a crossed network of ripples merging each other at pillar centers is observed. 
This different ripple pattern (ie. a second set of parallel ripples) is most probably attributed to the fabrication process which involve the \textit{in situ} formation of pillar array and suspension after transfer by selectively dissolving the underlying substrate.
In contrast, our fabrication process relies on the interaction of the free graphene layer in a fluidic environment with the prefabricated pillar array and avoids contamination or alteration of the graphene since it is transferred at the last step and not exposed to electron beam (see \ref{figS1} Supp. Info.). 
In our studies, we mainly focused on sharp pillars with variable lattice parameter $a$.


\section{Study of ripple domains formation and transition towards suspended graphene.}
\begin{figure*}[htbp]
\begin{center}
		\includegraphics[width=\textwidth]{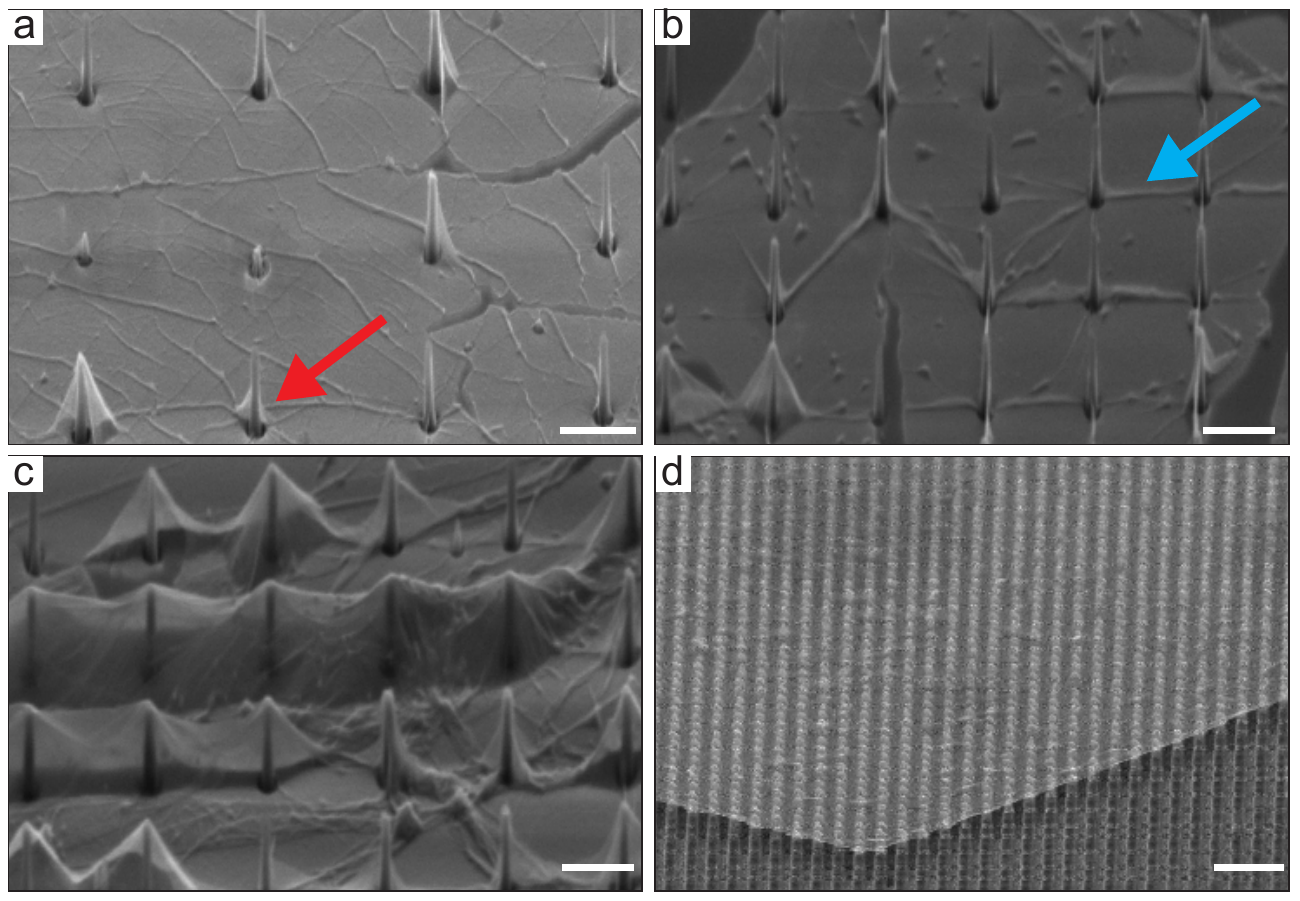} 
		\caption{\textbf{Graphene sheets deposited on corrugated substrates with increasing pillar density.} Series of SEM micrographs showing the behavior of transferred graphene membrane for increasing density of 270-nm-height silicon pillars. From \textbf{a} to \textbf{d}, the spacing between pillars is respectively equal to 2.3, 1.5, 1.4, and 0.25 $\mu $m.  For sharp and low-density-packed pillar arrays (\textbf{a}), ripples do not join neighboring pillar but rather show a preferential direction presumably reminiscent of the copper surface step edges on which the graphene has grown (see graph 8). Orientational order of ripples along the symmetry axes of the pillar network starts to appear for 1.5 $\mu$m pitch (\textbf{b}). At about the same density, partial suspension along symmetry lines could be observed (\textbf{c}), while for 250 nm spacing between pillars (i.e. for an aspect ratio about 1),  (\textbf{d}) the membrane becomes fully suspended in between pillars. Scale bars represent 2 $\mu$m.}\label{fig2} 
	     \end{center}
\end{figure*}

\ref{fig2} shows the graphene layer deposited onto a square lattice of SiO$_2$ nano-pillars with a large lattice spacing $a$ (typically $a\sim 1\mu$m). 
In such cases, the graphene sheet always fully collapse on its entire surface, forming a conformal capping layer on the corrugated sample with many frowns called "ripples" that will be further analyzed using SEM and Raman spectroscopy. 
The graphene ripples linking pillars are clearly visible, reminiscent to what can be observed on a cloth covering a non-flat surface, and its ordering becomes more apparent as the pillar array density increases. 
It is worth noting that depending on the array parameters (pillars height and spacing, \ref{fig2}a) the graphene hugs the pillars tight or hangs more loosely around the pillars in other cases (\ref{fig2}b-d), leading to partially suspended features.
In \ref{fig2}a-b, tears in the graphene are visible, appearing as straights dark stripes and are attributed to  correspond to grain boundaries preexisting in the CVD grown graphene membrane.
Different morphologies of ripple formations are observed experimentally: i) the graphene exactly coats the pillars in a conformal fashion  (cf. \ref{fig2}a), or ii) is locally suspended around the pillars, leading to a tent-like feature (cf. \ref{fig2}b-d).  In some cases, for instance as indicated by the arrow in \ref{fig2}b, the graphene ripples tend to be oriented parallel to the direction of the underlying pillar array (in such case, linking 1$^{st}$ nearest neighbors).
Experimentally, this collective behavior occurs if the tip of the pillar is sufficiently small compared to the curvature radius of the typical ripples observed (\textit{ie.} $\sim$ 42 nm, cf. \ref{fig4}b) and only in the same graphene grain boundary.
Other cases are possible, for example, when the inter-pillar distance is too large, the ripples from a pillar extend radially and point towards various directions (as indicated by the red arrow (\ref{fig2}a)) 

\begin{figure}[htbp]
\begin{center}
		\includegraphics[width=0.5\textwidth]{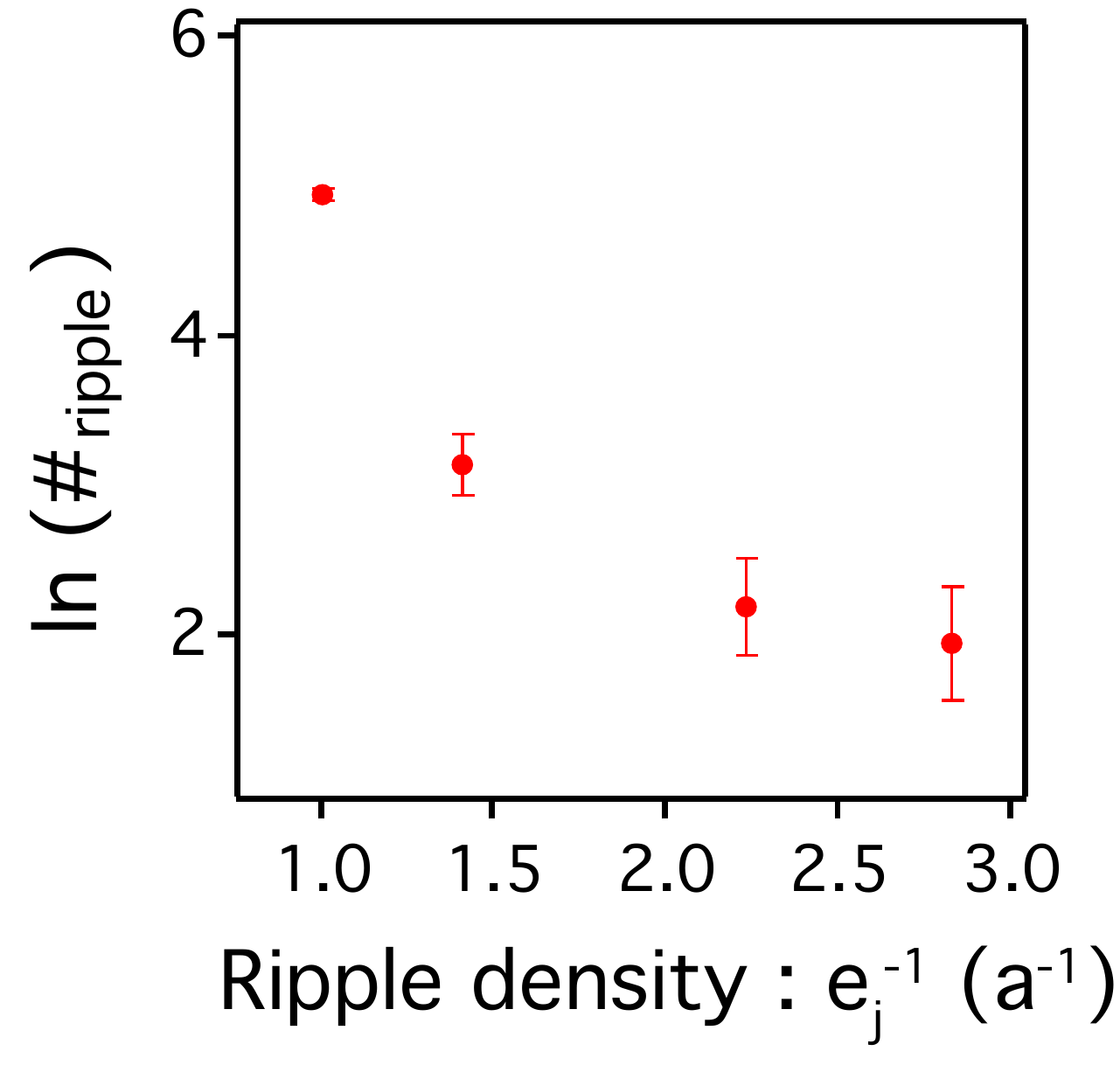}
	     \caption{\textbf{Statistical analysis of the distribution of ripples in graphene membrane. }\textbf{a}: Graphene ripples distribution as a function of the graphene ripples density $e_j^{-1}$ for different geometrical configurations (1$^{st}$, 2$^{nd}$, 3$^{rd}$ neighbor, etc). The notation $e_j$ is introduced at the \ref{eq:rho} (Supp. Info.). The experimental data have been extracted from a single square lattice (parameter $a=1.5 \ \mu$m). }\label{fig6B}
\end{center}

\end{figure}

\ref{fig6B} presents the occurrence of ripples linking 1$^{st}$, 2$^{nd}$, 3$^{rd}$ and 4$^{th}$ neighbors.
The closer neighbor configuration (ie. lower ripple density) is clearly dominant. 
These observations raise questions about the formation of graphene ripples, and their resulting geometrical configuration. 
In particular, on has to understand what is the driving force which lead graphene ripples to be aligned along high symmetry axes of the square lattice of nano-pillars.
When graphene is deposited onto the patterned substrate, its area is smaller than the area of the patterned specific surface of SiO$_2$ because of the 3D character of the nano-pillars. 
In other terms, there is an topological mismatch between two surfaces as an unstrained graphene membrane cannot fit the substrate exactly. There are two competing interactions are at work: i) the sum of all attractive interactions (Van der Waals, electrostatic, etc) which tend to force graphene to collapse onto the substrate and ii) the repulsion between $\pi$ orbitals of graphene which causes internal rigidity of the graphene sheet forcing it to remain as flat as possible\cite{Fasolino2007}.

In order to describe the competition between these antagonist interactions, we note $\mathcal{E}_{c}$ the energy density for the attractive interactions and $E_{r}$ the energy needed to create a graphene ripple. 
 Following the notation introduced by Tersoff \cite{Tersoff1992}, we consider a graphene ripple as a half cylinder, $E_{r} = \frac{c_0}{R^2}S_{0}$, where $c_0$ is an elastic constant for curvature out of the plane\cite{Zhu2012} ($c_0 =$ 1.4eV), $R$ the ripple radius (see Supp. Info.), and  $S_{0}=2\pi L R$ the surface of a cylinder (a ripple is viewed as two half cylinders of opposite curvature). 
In first approximation, we simplify the integral $\int \mathcal{E}_c(\vec{r}) d^2r$ as $S_c\mathcal{E}_c$.
A simple equilibrium condition can be written as :
\begin{equation}
\Delta E = S_{c}\mathcal{E}_c -E_{r} N_{r}= 0
\label{eq:bilan}
\end{equation}
 where $N_{r}$ is the number of graphene ripple contained in a surface $L^2$, and $S_{c}$ is the surface of graphene which is in contact with the substrate.
Regarding \ref{eq:bilan}, if $\Delta E >0$, the energy cost to bend graphene remains smaller compared to the total attraction energy. 
In that case, the transferred graphene membrane collapses and fits the substrate except at some particular positions, forming 1D ripples.
If $\Delta E <0$, the energy cost to bend graphene is now higher than the total attraction energy, and the transferred graphene membrane stays flat, resting fakir-like on top of the nano-pillars.

To connect the \ref{eq:bilan} to our experimental parameter $a$, we introduce the ripple density $\frac{a}{e_j}$, where $e_j$ is the characteristic distance between two parallel 1D graphene ripples (see supp. info.).
Therefore, the number of ripples of size $L$ in a given surface $L^2$ is $N_r=\frac{L}{e_j}$. 
Meanwhile, the surface of graphene in contact with the substrate is $S_c = e_jL-S_{susp}$, where $S_{susp}$ is the fraction of suspended graphene at the pillar edge and at the ripple.
Introducing these notations in \ref{eq:bilan}, we obtain a critical value of $a^*$ satisfying $\Delta E =0$ (cf. Supp. Info.).
This value of $a^*$ separates the two regimes of fully suspended graphene from collapsed and rippled graphene.
Interestingly, expression of $a^*$ derived in \ref{eq:astar} (see Supp. Info.) qualitatively explains the two main observations of our work : i) the reduction of $a$ leads to a full suspension of graphene when $a<a^*$, and ii) ripple orientation is statistically in favor of the nearest neighbor configuration found in the low ripple density regime.
In addition, the critical parameter $a^*$ also show dependance in $\mathcal{E}_c$ which is related to the physisorption properties of the substrate.
Thus, this result has a crucial importance in order to engineer the corrugated substrate to pre-determine the transferred graphene properties. These properties are governed by the generated stress and doping in the two different regimes ($a<a^*$ or $a>a^*$). Both stress and doping are now probed by Raman spectroscopy. 

\section{Raman analysis of collapsed and suspended membranes : controlled stress }
\begin{figure}[htbp]
\begin{center}
		\includegraphics[width=\textwidth]{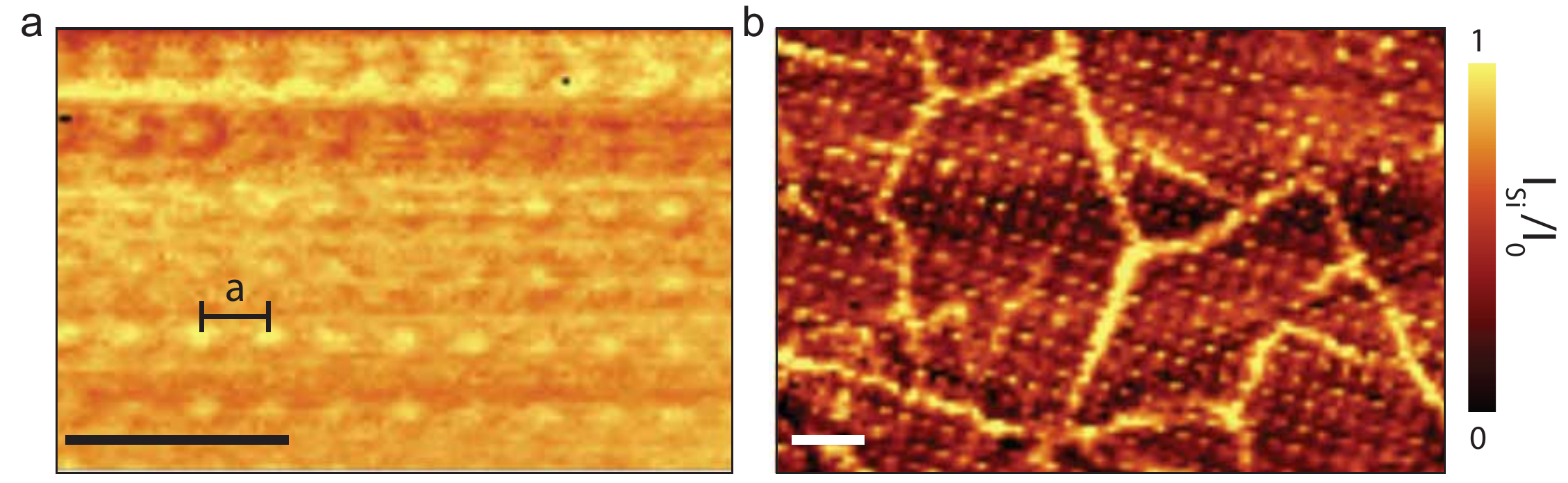}
	     \caption{\textbf{Raman map of TO-silicon peak intensity.} Raman map of the intensity of the Si-TO mode at 521 cm$^{-1}$ before (\textbf{a}) and after (\textbf{b}) the graphene transfer. The configuration of the transferred graphene is likely to be similar to the case presented in \ref{fig2}b-c. Scale bars represent $3\ \mu$m.}\label{fig3}
\end{center}
\end{figure}

In order to have an estimation of the generated stress depending on the different configurations, we use Raman spectroscopy which is a powerful tool to reach this goal and allow us also to approach the critical value $a^*$. 
First of all, we need to precisely locate the position of the nano-pillars before analyzing the Raman response of deposited graphene.
For this purpose, we investigate the silicon TO Raman active peak (Si-TO) at 520.7 cm$^{-1}$ \cite{Jr1967}.
The Raman spatial map in \ref{fig3}a shows that the intensity of Si-TO peak follows the nano-pillars periodicity. 
We find that the top of a nano-pillar exhibits higher Raman intensity I$_{Si}$ than the bottom plane.
This phenomenon is explained by optical interference enhancement\cite{Skulason2010,Blake2007,Connell1980,Reserbat-plantey2013,Reserbat-Plantey2012}.
We consider the optical cavity made by a silicon mirror and a semi-transparent SiO$_2$ layer of thickness varying from 300 nm at the pillar base, to 560 nm at the pillar top. 
The height of the pillar is therefore approximately half the wavelength of the scattered light (\textit{ie.} 270 nm).
Due to optical interference between scattered beams, the collected Raman signal is modulated when the SiO$_2$ thickness varies by $\lambda/(2n)$ (where $n$ is the optical index of SiO$_2$).
Therefore, interference conditions are different from the top to the basis of one nano-pillar, which explain the modulation of the collected Raman scattered light along the substrate.
When graphene is deposited on top of the nano-pillars array, I$_{Si}$ still indicates the position of the nano-pillars (cf. \ref{fig3}b).
It is worth noting that the optical focus depth is about 700 nm, thus greater than the pillar height, excluding any defocusing effect in that I$_{Si}$ modulation.
Nevertheless, at a pillar position, optical conditions differ since the top part of the optical cavity is now the graphene layer, absorbing 2.3 \% of light and defining new optical interference conditions\cite{Nair2008}.
Note that \ref{fig3}b yields informations on the polycristallinity of the graphene layers, made up of grains of different sizes.
These grain boundaries are easily identified on the I$_{Si}$ Raman map because the silicon Raman signal is higher where there is no graphene.
In both cases, before and after graphene deposition, the frequency of the Si-TO peak (see Supp. Info.) does not vary along the nano-pillars array.
The analysis of Raman signal of the Si-TO peak is thus a good mean to determine the position of the nano-pillars and consequently is helpful for the interpretation of the graphene Raman response.

\begin{figure}[htbp]
\begin{center}
		\includegraphics[width=\textwidth]{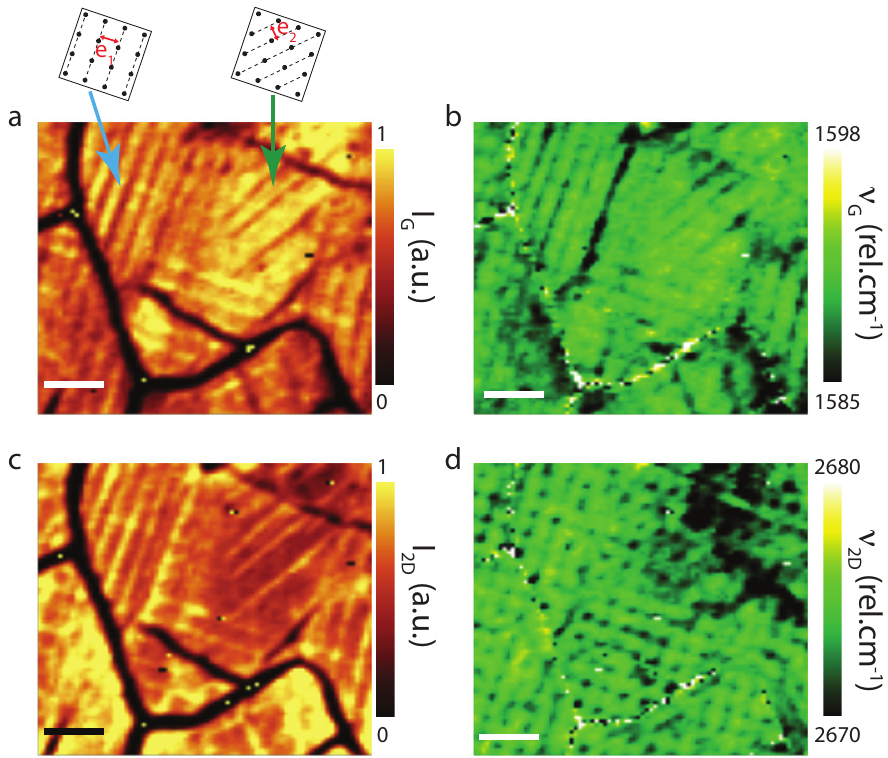}
	     \caption{\textbf{Strain domains in graphene deposited on SiO$_2$ nano-pillars. } \textbf{a-b}: Raman maps of the G band intensity (\textbf{a}) and frequency (\textbf{b}). The blue and green arrows show strain domains with 1$^{st}$ and 2$^{nd}$ nearest neighbors configuration, respectively (cf. insets where black dots symbolize the position of the nano-pillars and dashed lines represent the 1D graphene ripples. \textit{NB:} sketch scale is different from data scale). The distance $e_i$ between two consecutive ripples is represented in red. \textbf{c-d}: Raman maps of the 2D band intensity (\textbf{c}) and frequency (\textbf{d}). The configuration of the transferred graphene is likely to be similar to the case presented in \ref{fig2}b-c. Laser wavelength is 532 nm. Scale bars represent 3 $\mu$m.}\label{fig4}
\end{center}
\end{figure}

The Raman response of monolayer graphene always shows G and 2D peaks which are shifted in frequency in the presence of strain\cite{Frank2010,Frank2011a,Huang2010,Mohr2004,Yoon2011}.
For large strain ($\epsilon >$ 2 \%), it is possible to experimentally measure a mode splitting of the G peak, giving rise to G$^+$ and G$^-$ peak.
However, the Raman signature of graphene is also very sensitive to doping\cite{Ferrari2008} or thermal effects\cite{Calizo2007}.
Because of the bimodal dependence of both G and 2D band to strain and doping, it is rather complex to distinguish those two effects.
Nevertheless, correlations between the frequency of G and 2D bands give a clear signature of the importance of doping and strain \cite{Lee2012}. 
During this experiment, laser power is kept at 500 $\mu$W.$\mu$m$^{-2}$ in order to avoid heating effects, that are observed at higher laser power.  
Moreover, we carefully correlate the position of the nano-pillars with the position of the observed shift on the graphene Raman signature.

\ref{fig4} shows confocal Raman maps in the plane of the substrate.
The two arrows in \ref{fig4}a point to domains where Raman signatures of the G and 2D peaks are non-uniform and oriented along a single direction, forming parallel lines.
Knowing the exact position of the nano-pillars from the Si-TO peak, we assign the domain pointed to by the blue (green) arrow to be constituted by graphene ripples linking the 1$^{st}$ (2$^{nd}$) nearest neighbors.
The angle between the ripples axis on these two domains is equal to 45$^{\circ}$, in agreement with this assignment.
Graphene ripple lines are observed in the intensity signal of the Raman G and 2D bands, but also in the frequency mapping  - $\nu_G$ and $\nu_{2D}$ - of these Raman bands (see \ref{fig4}-b,d).
A downshift of $\Delta \nu_G =$ -2.8 cm$^{-1}$ from a ripple region to a flat one corresponds to a stretching of the graphene membrane of about 0.1 \%. 
It is worth noting that value may be underestimated since the laser spot is about 6 times larger than the ripple diameter (cf. \ref{fig6}-b), therefore the Raman signals of both flat unstrained graphene and highly strained graphene ripple are averaged. 
Even if $I_{2D}$ map exhibits ripples lines, the frequency of the 2D band - $\nu_{2D}$ - is mainly correlated to the nano-pillars position. 
\ref{fig4}d shows that $\nu_{2D}$ is minimum on top of each nano-pillar.
Previous experiments\cite{Berciaud2009} have shown a similar effect on the frequency of the 2D band of locally suspended graphene.
This frequency downshift can be attributed to the decrease of electrostatic interaction between graphene and the substrate, which occurs in regions where graphene is locally suspended as suggested by \ref{fig2}-b,c.
It is worth noting that such decrease of the 2D peak frequency can be attributed to the stress within the graphene sheet when it is pinned at the top of a pillar. 
Both effects, electrostatic doping and stress, can reasonably be considered at the pillar position. 
High spatial resolution Raman investigation, such as TERS, and cross polarization Raman analysis would give insights to discriminate between these two effects.

\begin{figure}[htbp]
\begin{center}
		\includegraphics[width=\textwidth]{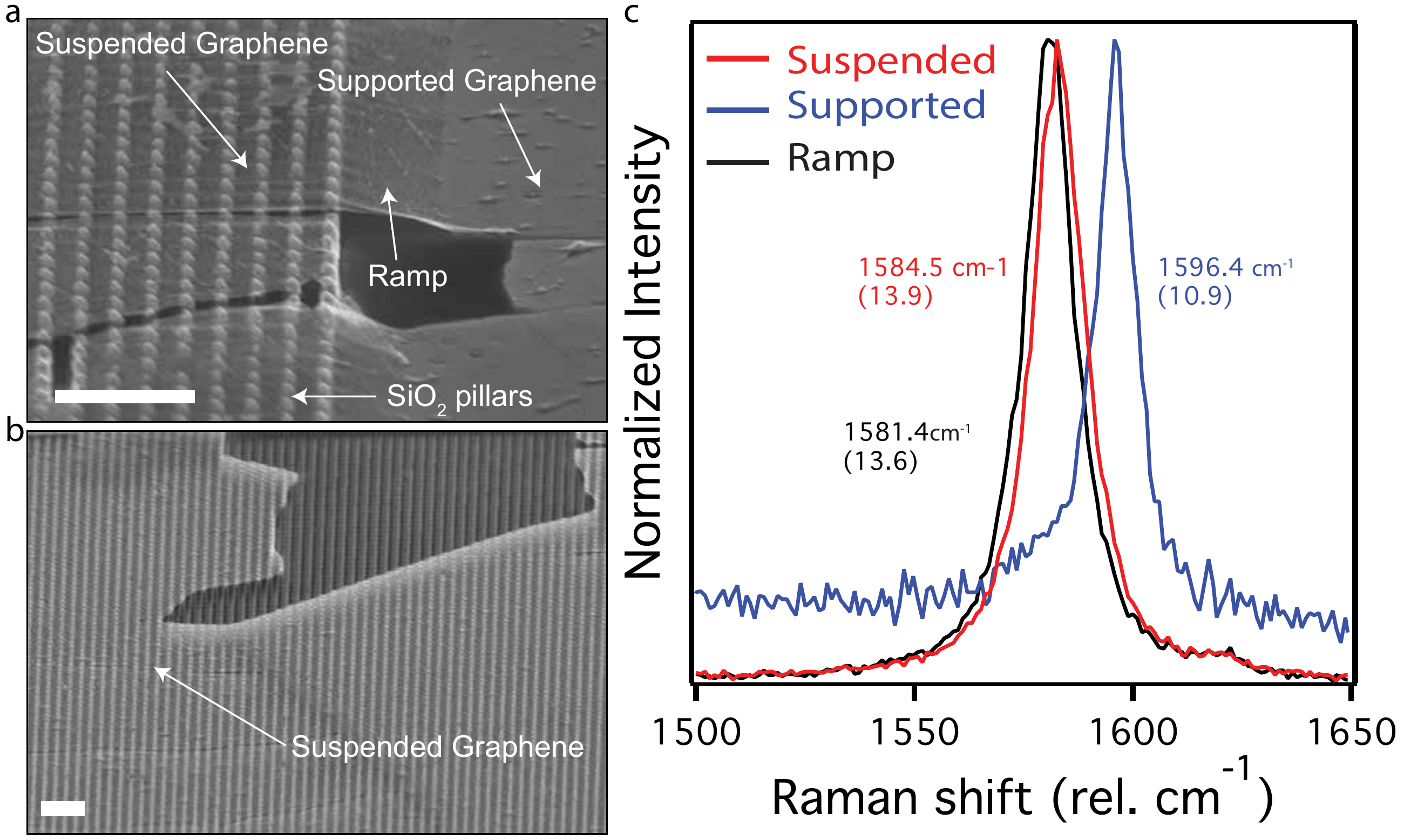}
	     \caption{\textbf{Full suspension of graphene on top of high-aspect ratio pillar arrays. a-b}: SEM micrographs of suspended graphene membrane on top of nano-pillars square lattice ($a=$ 250 nm). Scale bars represent 1$\mu$m. Note the presence of tears at graphene grain boundaries (a), which are consistent with previous Raman maps (cf. \ref{fig4}). \textbf{c}: Raman spectra (G band) of monolayer graphene suspended on top of nano-pillars square lattice (red line), collapsed on the substrate (blue line), and at suspended ramp region (dark line) for comparison. The frequency (width) of the G band is indicated on the graph.}\label{fig5}
\end{center}
\end{figure}

Until now, we have examined low pillar density (\textit{ie.} $a>1 \mu$m) case, in which graphene is lying on substrate and forms aligned ripples. 
However if the pillars lattice parameter $a\le a^*$, the system should not be considered in the intermediate regime where the attractive and repulsive interactions are almost equilibrated since the bending of graphene is no longer more favorable at that scale (cf. Supp. Info.). 
We observe that below a critical value $a\le a^*$, the deposition leads to fully suspended graphene over large areas (cf. \ref{fig5}-ab).
We have determined a upper and lower boundary for the value of the critical lattice parameter $a^*$:  $250$ nm $<a^*<$ 1 $\mu$m.
Raman spectra of the G band for suspended and supported graphene on a flat region (outside the nano-pillar lattice) are shown in \ref{fig5}-c. 
In the suspended case, the G band frequency shows a downshift of about  $\Delta \nu_{G} =$ -11.9 cm$^{-1}$ with respect to the supported case as well as a reduction in width ($\Delta \Gamma_{G} =$ -3 cm$^{-1}$).
Considering that the electrostatic influence of the substrate (\textit{ie}: charge impurities) strongly weakens when graphene is suspended, this softening observed on the G band is interpreted as a consequence of the decrease of charge transfer between the graphene and the substrate\cite{Berciaud2009}.
Analysis of the 2D band (cf. Supp. Info.) also confirm that graphene is less doped in the suspended case than in the supported one.
According to ref\cite{Lee2012} the increase in carrier density between the suspended case and the supported one is about 8 $.10^{12} cm^{-2}$. 
This confirms the reduction of doping for such macroscopic suspended graphene sheet.
Nevertheless, contribution of residual strain due to pinning graphene at the top of nano-pillars array would also downshift the G band frequency.
By comparison between the strained graphene on the ramp and the low-doped suspended graphene, a rough estimation of this strain can be obtained, and is about 0.1 \%, which is quite similar to the strain value extracted at a pillar position in the case where $a>a^*$. 
Note that this estimation is an average value because our spatial resolution is bigger than the inter-pillar distance. 
Moreover, no G band splitting has been observed in the suspended region (cf. \ref{figS7}), as it is expected in case of strong uniaxial stress, which confirm that stress contribution to the change on the Raman response is not major in that particular case. 
Note that doping alone cannot explain all the Raman feature (G and 2D) observed in the supported region.
It is therefore likely that the suspended graphene is less strained than the supported one. 
To summarize, when macroscopic graphene sheet is suspended on top of the nano-pillars, both stress and doping are decreased in comparison with supported graphene.
This might be an important feature for future integration of high mobility electronic devices.

So far, we have examined a square lattice of nano-pillars.
Considering now another type of pillars lattice, for example a random lattice, the ripple propagation should then be impossible because of absence of high symmetry axis to maintain the ripple propagation. 
In such case, the average distance between $N$ pillars distributed over an area $S$ would be $a_r = \sqrt{S/N}$ and the critical value to get fully suspended graphene would be lower than the square lattice case $a_r^* < a^*$.
A recent study\cite{Yamamoto2012} highlights the effect of pillars density (made from randomly deposited nanoparticles) on graphene ripples formation. 
These authors show AFM measurements leading to a critical pillars density for which graphene ripples form a percolating network. 
In addition to the present work, this is therefore a first step towards large areas of fully suspended graphene.
Devices made of macroscale suspended graphene are of interest both for fundamentals investigations (role of periodic potential created by the pillars, collective low energy vibration mode, ...) and for applied science (high mobility transparent electrodes, batch fabrication of mechanical resonators, ...)


\section{Conclusion}
\bigskip
In conclusion, using a set of prefabricated substrates with pillar arrays of variable aspect ratio, we have provided a platform to study the formation of suspended graphene membranes over tens of micrometers and the transition from this suspended state to a collapsed one which exhibit organized domains of parallel ripples joining the pillars. 
Depending of the array geometry and pitch, graphene films can tightly coat the surface, partially collapse, or lay, "fakir-like", suspended for an array parameter below 1$\mu$m (pillar height of 260 nm).
Different cases allow to illustrates the competition between adhesion and membrane rigidity. 
Collapsed films display set of parallel ripples organized in domains, thus forming strain domains of different configurations. 
These ripples are oriented along high symmetry axes of the pillars lattice.
Such collective behavior is qualitatively described taking into account the ripples density and the corresponding bending energy.
Stress domains are then observed by Raman spectroscopy mapping and correspond to parallel ripples regions. 
Typical stress at the graphene ripple is about 1 GPa.
Raman spectroscopy appears as a reliable and non invasive investigation tool to quantify stress, discriminate strained domains and identify order in the strain organization..
In addition to controlling the stress of a graphene once transferred onto a substrate (control of ripple formation, and local strain), we have shown that, by increasing the aspect ratio of the pillar, a transition towards a macroscale suspended graphene membrane takes place. In that latter case, the interaction with the substrate is becoming minimal and offer a promising way to test the influence of suspended graphene which periodic substrate interaction.  

\acknowledgments
This work was supported by the Agence Nationale de la Recherche (ANR projects : MolNanoSpin, Supergraph, Allucinan and Trico), European Research Council (ERC advanced grant no. 226558), the Nanosciences Foundation of Grenoble and Region Rhône-Alpes and the Graphene Flagship.
Authors thanks Edgar Bonet and Joël Moser for stimulating discussions.

\section{Supplementary Material}

\section{Fabrication techniques}

Graphene is grown by CVD process on copper. 
In short words, 25 $\mu$m thick Cu foil is loaded in a quartz tube under 1 mbar total pressure, and 1000 $^{\circ}$C annealing for 1 hour was applied. 
Keeping the same temperature, graphene growth was performed with 2 sccm CH$_4$ and 1000 sccm H$_2$, while the total pressure was changed into 25 mbar. 
After 10 min growth, H$_2$ and CH$_4$ is shut down immediately, instead 500 sccm Ar is injected, and the setup is cooled down to room temperature in 3 hours. 
The result graphene are in hexagonal shape. 
Note that there are wrinkles or ripples within the graphene layer, due to the mismatch of thermal expansion coefficient of Cu and graphene (corrugated graphene was inevitably folded into wrinkles when transferred onto substrate), as previously reported\cite{Ruoff}. 
Graphene shown in \ref{fig2} b,d comes from CVD graphene supermarket (\url{http://graphene-supermarket.com/}).

\begin{figure}[htbp]
\begin{center}
		\includegraphics[width=0.5\textwidth]{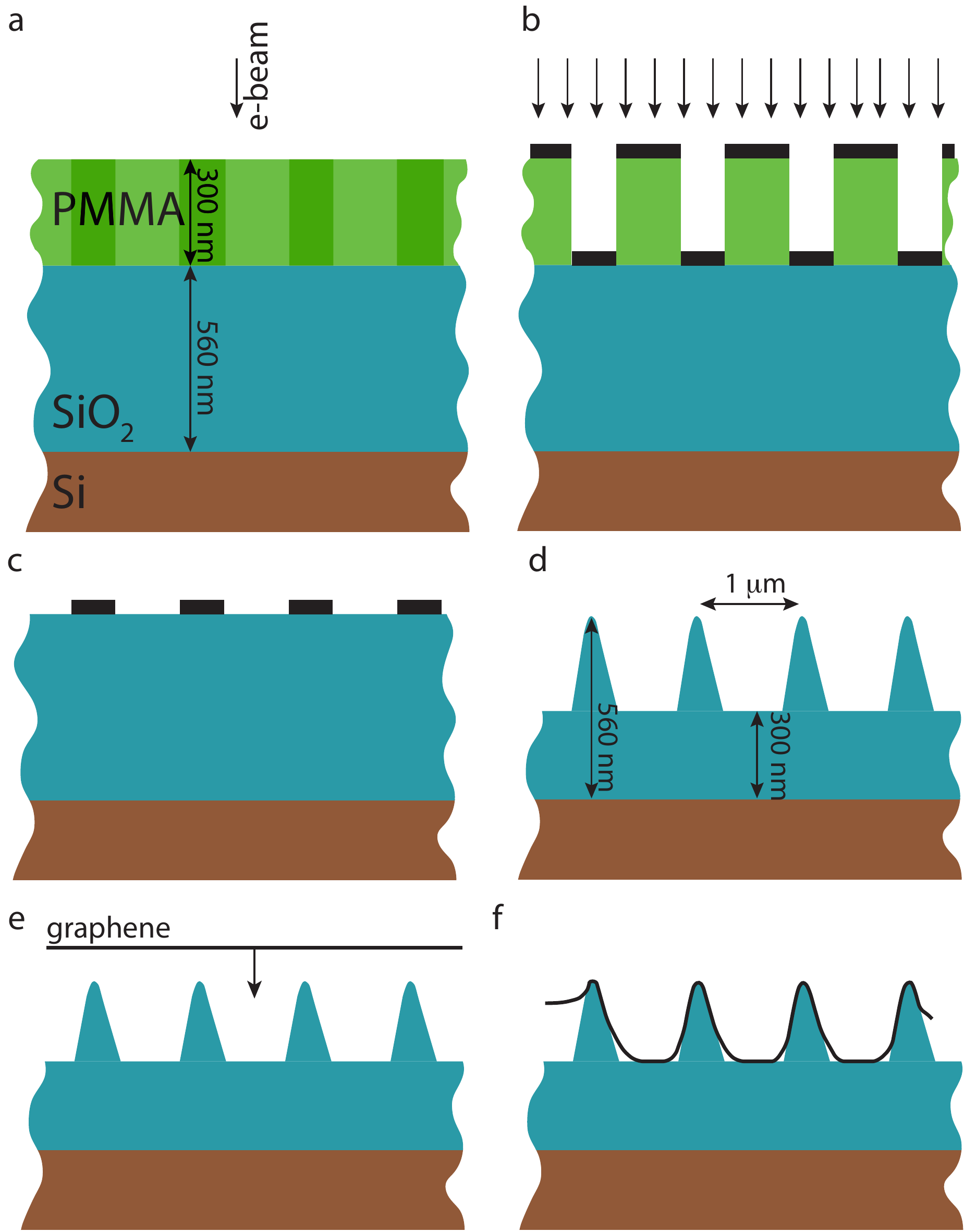}
	     \caption{\textbf{Fabrication process of the sample.} \textbf{a}: electron-beam lithography of pilars. \textbf{b}: 50 nm Ni or Al deposition. \textbf{c}: RIE etching (SF$_6$/CHF$_3$).\textbf{d}: Metal (Ni) dissolution in HNO$_3$ to obtain nano-pillars of SiO$_2$. \textbf{e}: Transfer of CVD graphene onto the nano-pillars array. \textbf{f}: graphene layer lying on top of nano-pillars.}\label{figS1}
\end{center}
\end{figure}

Nano-pillars substrate fabrication is schematically described in \ref{figS1}.
One 300 nm-thick layer of PMMA is spin-coated onto a oxidized silicon wafer with 560 nm of SiO$_2$. 
The nano-pillars pattern is designed by electron-beam lithography. 
After development, 50 nm of Ni or Al metal are deposited, which constitute the mask to create the nano-pillars by plasma etching (RIE) of the silicon oxide. Al and Ni masks lead to different apex shape and sharpness as Al is partially etched during RIE while Ni is not. 
The remaining metal (Ni) is dissolved in HNO3 solution. 

The transfer of graphene onto nano-pillars starts by spin-coating the graphene on Cu with PMMA. 
Graphene on the back-side of the Cu foil was etched by oxygen plasma and then the PMMA/graphene/Cu stack was floated on Ammonium persulphate ((NH$_4$)$_2$S$_2$O$_8$) diluted with DI water (0.02 mg/ml) for 24 hours. 
Once all Cu is removed, the sample PMMA/Graphene is carefully washed  in fresh DI water for at least 10 times. 
The transfer is then realized by slowly picking-up from below the PMMA/Graphene layer with clean nano-pillar substrate, followed by a natural drying in air for one hour. 
To increase the chance of sticking onto substrate, the whole sample was soft-baked on a hotplate at 120$^{\circ}$C for 5 min. 
The PMMA layer is eventually removed with acetone and dried from an IPA rinse.

\begin{figure}[htbp]
\begin{center}
		\includegraphics[width=0.7\textwidth]{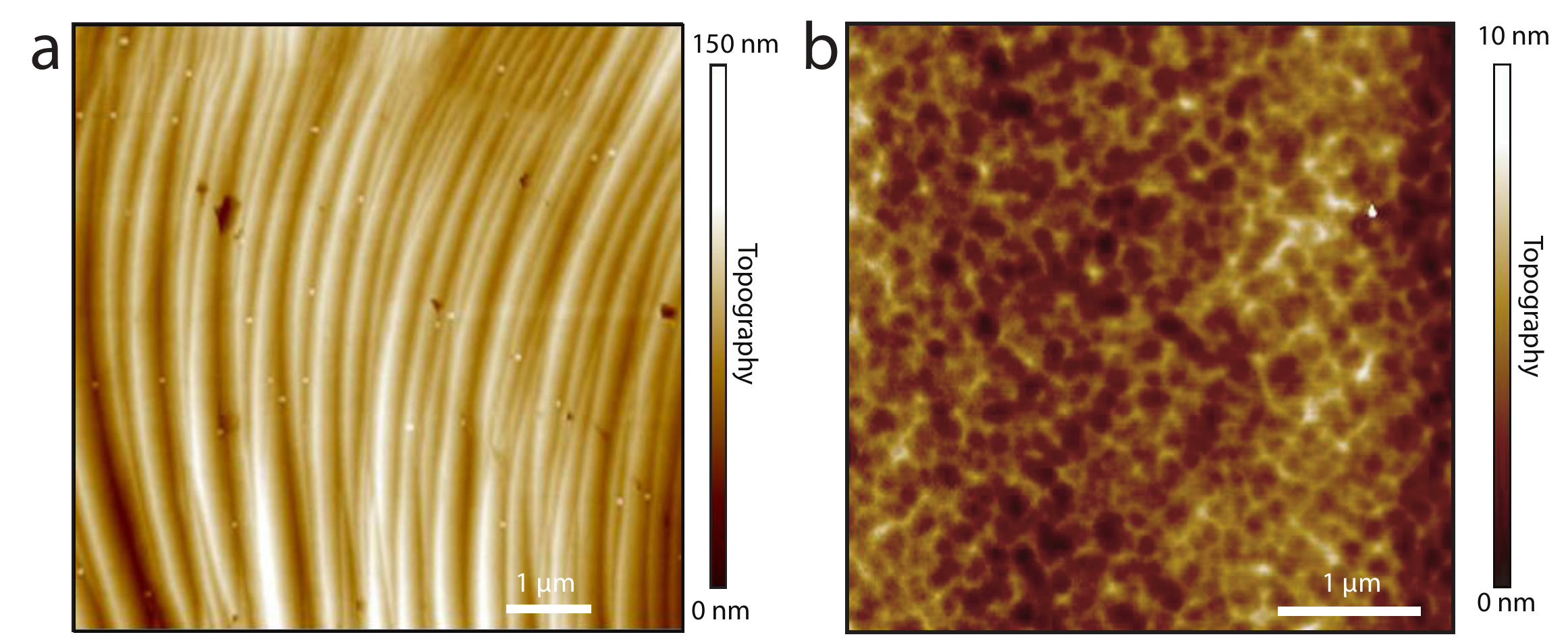}
	     \caption{\textbf{AFM topography of graphene on PMMA, just before the transfer.}\textbf{a-b} : topography mirographies of graphene attached on PMMA layer, corresponding of the snapshot of step e in \ref{fig1}. Graphene wrinkles are observed, probably due to the finger print of Cu-terrasses. Orientation of those wrinkles may vary from one end to the other of the chip. }\label{figAFM}
\end{center}
\end{figure}

\section{Raman spectrometer setup}
The setup consists of a confocal microscope with a 320 nm spot size for $\lambda_{laser}$ = 532 nm.
Confocality of the system is ensured by a 50 $\mu$m optical fiber for both injection and collection of light.
The elastically scattered light from the sample is filtered out by an edge filter, while the inelastically scattered light is collected and sent to a spectrometer with resolution less than 0.9 cm$^{-1}$.
Spectrum acquisition is performed by a CCD camera, cooled down to -65 $^{\circ}$C by Peltier cooling.
A typical Raman spectrum is acquired in 1-10s.
To avoid laser heating, laser power is kept below $P_{laser} = 0.5 \ \rm{mW.\mu m^{-2}}$.
The Raman spectrometer (WITec alpha 500) is equipped with a piezoelectrical stage, allowing to make 3D confocal maps of the sample.
\begin{figure}[htbp]
\begin{center}
		\includegraphics[width=\textwidth]{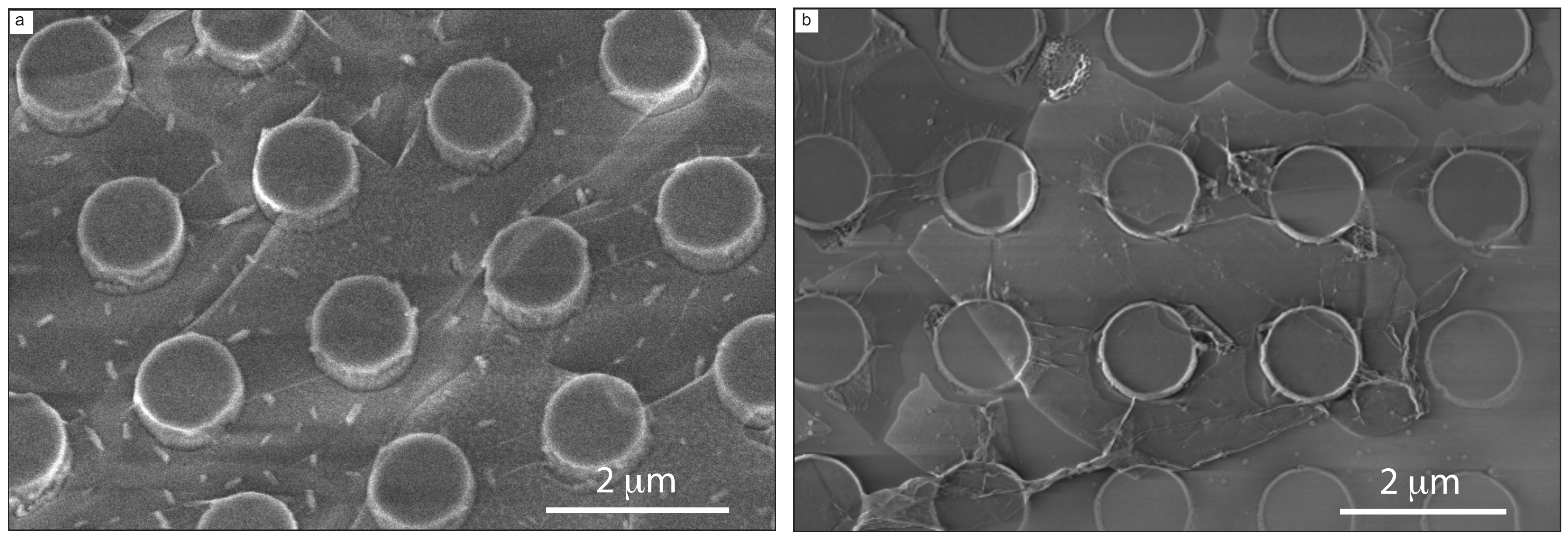}
	     \caption{\textbf{Role of the nano-pillar tip.} \textbf{a-b}: SEM micrographs of graphene transferred on top of large area nano-pillar (tip area $\sim 1 \mu$m$^{2}$, tip radius $\sim 0.55 \mu$m. No collective behavior of ripple propagation is observed. However, ripples point towards multiple directions from one pillar, confirming the hypothesis about destruction of ripple coherence (and propagation) when nano-pillar tip is larger than ripple natural radius ($\sim 42$ nm).}\label{fig:gros_piliers}
\end{center}
\end{figure}
\section{Neighbors indexing}
Each set of parallel ripple is defined by $m_j$ and $n_j$ which are integer indexes for the $j^{th}$ nearest neighbor nano-pillar configuration. 
For instance, $(m_j, n_j) = (0,1)$ represents the first neighbor configuration ($e_1 = a$), and $(1,1)$ the second neighbor configuration ($e_2 = a/\sqrt{2}$). 
It is possible to derive the following expression for the square lattice: 
\begin{equation}
e_j = \frac{a}{\sqrt{m_j^2+n_j^2}},
\label{eq:rho}
\end{equation}
These indexes follow two selection rules: i) $e_j > e_{j+1}$, and ii) $\frac{n_{j_1}}{m_{j_1}} \ne \frac{n_{j_2}}{m_{j_2}} \ \ \ \forall
\{(m_{j},n_{j})\} \in  \{ m_j^2+n_j^2 = A \}$ (where $A$ is an integer).
The first condition imposes that the $j^{th}$ nearest neighbor is always closer to the initial position than the $(j+1)^{th}$ one.
The second condition avoids double counting of neighbors and includes the degeneracy of the ripples configuration. 
As an example (1,1) and (2,2) both correspond to the same ripple while (3,1) and (1,3) are characterized by the same density of ripples. 
Note that the degeneracy of the ripple configuration depends on $j$ (as an example 1$^{st}$ and 2$^{nd}$ are doubly degenerate, while 3$^{rd}$ nearest neighbor configuration is 4-times degenerate).

\section{Derivation of the critical parameter $a^*$}
This parameter corresponds to the critical  value of the inter pillar distance  $a^*$ that leads to a transition from fully suspended to partially collapsed graphene.

\begin{figure}[htbp]
\begin{center}
		\includegraphics[width=\textwidth]{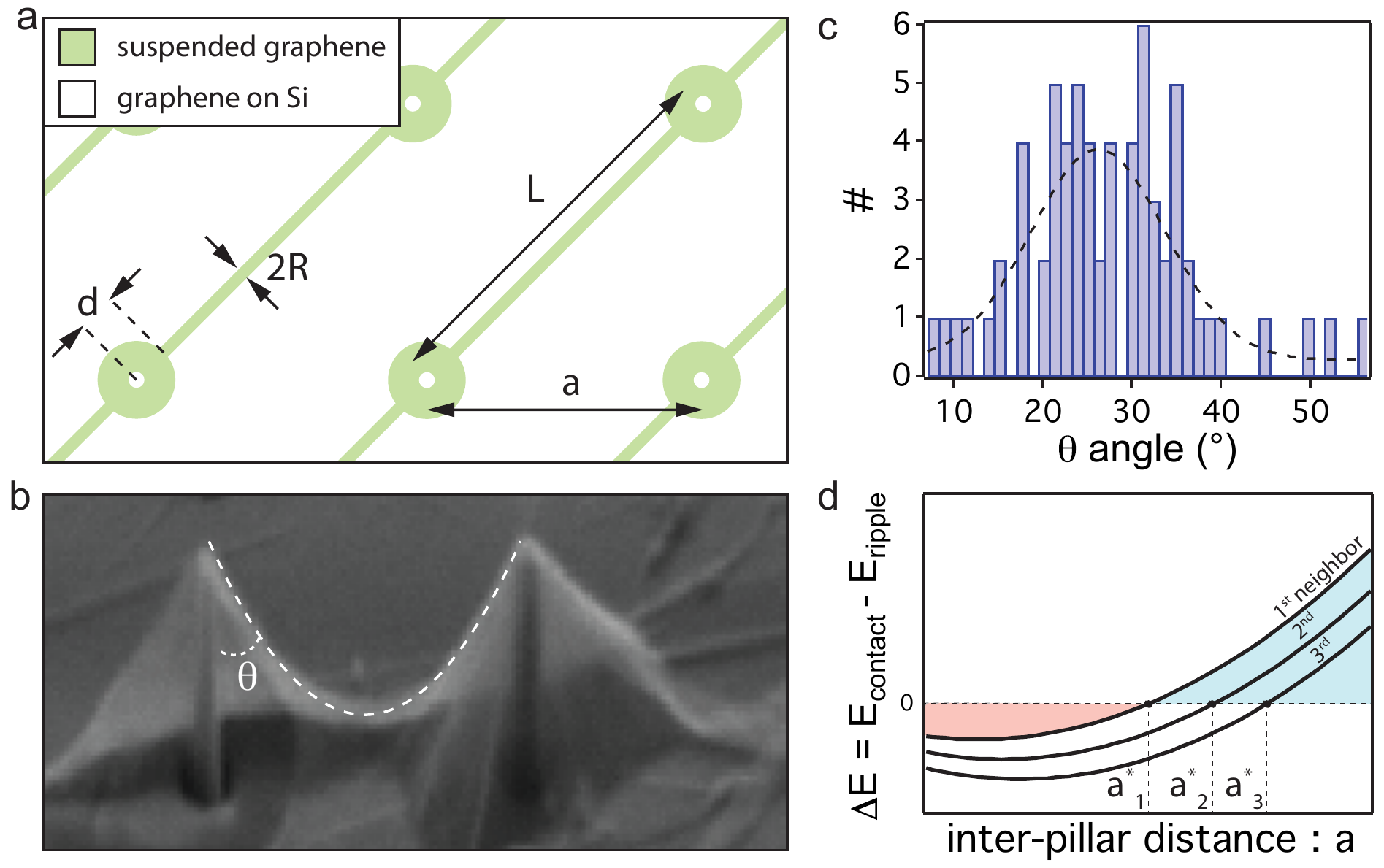}
	     \caption{\textbf{Geometry of the graphene ripple.}\textbf{a}: Top-view sketch of graphene transferred on nano-pillars and showing ripples. Green regions represent the surface where graphene is not in contact with the substrate. Around each nano-pillar, the graphene remains suspended and makes a tent-like conical shape.\textbf{b}: SEM micrograph of a typical case of graphene around a nano-pillar. Locally, suspended graphene can be modeled as a cone having an angle $\theta$. \textbf{c}: Statistical distribution of $\theta$ measured on 66 ripples. Mean value is $\theta = 26^{\circ}$. \textbf{d}: qualitative evolution of $\Delta E$ with the parameter $a$ for different neighbors. We observe that $a^*_{j+1}>a^*_j$. For the first neighbor configuration, if $a>a^*_1$, graphene shows ripples and stays in contact with the substrate (blue region). If $a<a^*_1$, graphene is fully suspended (red region).}\label{figS9}
\end{center}
\end{figure}

Starting from \ref{eq:bilan}:
\begin{equation}
\begin{array}{l}
\Delta E = S_{c}\mathcal{E}_c - E_{r} N_{r}= 0,\\
(Le_j - 2LR - \pi h^2\tan^2\theta)\mathcal{E}_c - E_{r} \frac{L}{e_j} = 0
\end{array}
\end{equation}
\label{eq:bilan2}
Where $R$ is the radius of a ripple which has been experimentally determined by SEM (cf. \ref{fig6}-b), $h$ the pillar height and $\theta$ the angle between the graphene and the pillar as depicted in \ref{figS9}. 
The surface $S_c = e_jL-S_{susp}$ correspond to the surface of graphene in contact with the substrate (cf. \ref{figS9}a). 
Therefore, it is equal to difference of the surface separating two consecutive ripples ($e_jL$) and the total suspended area $S_{sups} = 2LR + \pi h^2 \tan^2\theta$.
To estimate $S_{sups}$, we take into account the fraction of graphene that is not in contact with the substrate along a ripple ($2LR$), and the fraction of suspended graphene (conical shape with angle $\theta$) hanging around a pillar of height $h$.
In addition, it is worth noting that the term $\mathcal{E}_c$ varies along the position of graphene as the distance to the substrate may change locally \cite{Katsnelson2012}.

The critical parameter $a^*$ corresponds to the case where $\Delta E = 0$.
Considering the  lowest ripple density (ie. first neighbor configuration), $e_j = a$ and $L=a$. 
The previous equation can be rewritten as :
\begin{equation}
a^2 - 2aR - \left ( \pi h^2 \tan^2 \theta + \frac{E_r}{\mathcal{E}_c} \right ) = 0
\label{eq:polynom}
\end{equation}
Only the positive solution of \ref{eq:polynom} has physical meaning :
\begin{equation}
a^*= R + \frac{\sqrt{(2R)^2+4\pi h^2 \tan^2 \theta + 4\frac{E_r}{\mathcal{E}_c} }}{2}
\label{eq:astar}
\end{equation}
This result qualitatively predicts: i) the dependance of $a^*$ with the pillar height, ii) the dependance with the  ripple width, iii) full suspension of graphene for $a<a^*$ and iv) predominance of first neighbor configuration.
On can derive four qualitative results. 
\begin{itemize}
\item As $h$ increases, the distance separating graphene from substrate increases as well. This lead to a decrease of the total attraction energy between graphene and substrate, and therefore, $a^*$ would increase.
\item Wider ripples ($R$ large) can be explained as a consequence of a high bending energy. This is in favor of fully suspended graphene, without ripples. Therefore, for a given set of parameters, $a^*$ increases with $R$, as predicted in \ref{eq:astar}.
\item Experimentally, when $a$ is large, we observe ripple formation. Decreasing $a$ leads to a structural transition where graphene remains fully suspended. This suggest existence of a critical parameter $a^*$. We have shown that equation \ref{eq:bilan2} can be written as a polynom of $a$ having a positive quadratic term. Therefore, as $a$ is decreasing, there is a mathematical solution for $\Delta E =0$ at $a=a^*$. This value of $a^*$ separates the regime where $\Delta E>0$ (graphene ripples), from the one where $\Delta E <0$ (suspended graphene).
\item This last argument is less trivial. From equations \ref{eq:bilan2} to \ref{eq:polynom}, we have made an approximation considering only nearest neighbor configurations. Full derivation, including the indexes $n_j$ and $m_j$ leads to another critical value $a^*_j$. We find that $a^*_{j+1}>a^*_j$. This suggests that for $a \in  [a^*_1 ; a^*_2 ... a^*_j]$, the system prefers the most favorable configuration : $a^*_1$ which correspond to the first neighbors.
\end{itemize}
Experimentally, we can determine a set of parameters such as $a^* = 250$nm, $R=42$nm (cf. \ref{fig6}b), $\theta = 26^{\circ}$ (cf. \ref{figS9}c), $h=260$ nm (cf. \ref{fig1}) and we can calculate $E_{r} = \frac{c_0}{R^2}S_{0}$, where $c_0$ is an elastic constant for curvature out of the plane\cite{Zhu2012} ($c_0 =$ 1.4eV).
We obtain an adhesion energy around 5 mJ.m$^{-2}$, which is two orders of magnitude smaller than the value of 450 mJ.m$^{-2}$ measured by Koenig \textit{et al}\cite{Koenig2011} for monolayer graphene on SiO$_2$. 
This discrepancy could be explained by our overestimation of the ripple radius using SEM contrast images (cf. \ref{fig6}b. 

Generalization of \ref{eq:astar} taking into account other neighbors leads to: 
\begin{equation}
a^*_j= d_jR + \frac{\sqrt{d_j(2R)^2+4\pi h^2 \tan^2 \theta + 4\frac{E_rd_j^2}{\mathcal{E}_c} }}{2}
\label{eq:astar_gene}
\end{equation}
\ref{figS9}d shows the qualitative evolution of the critical parameter $a^*_j$ as a function of the neighbors.

\section{Statistical Modeling of ripples domain formation}
 In order to gain insights about the ripple formation and the fact that low ripple densities are dominant, we have developed the following toy model. 
In such model, we make the following hypothesis : i) a ripple propagates along one direction and is parallel to another ripple located at a distance $e_j$, ii) the energy to create a ripple costs $E_r$, iii) we only consider a system of a fixed number of ripples  $N$ , and therefore the contribution for attractive interaction with the substrate is a constant, and iv) the lattice parameter is  above critical value: $a>a^*$. 
For a system containing $N$ ripples, the total energy is then $E_T = E_{r}N$.
We now consider a system of size $L^2$, containing $N$ ripples of length $L$. 
It is worth noting that the ripples inside the system of size $L^2$ are not independent as we consider a set of parallel ripples. Therefore, the two indexes $(n_j,m_j)$ govern the configuration state. 
$N$ is given by the length of the system divided by the inter-ripple distance, ie. $N = \frac{L\sqrt{n_j^2+m_j^2}}{a}$.
Combining the precedent equations, the total energy of such system is : 
\begin{equation}
E_T = \frac{E_{r}L}{a}\sqrt{n_j^2+m_j^2} = \frac{E_rL}{e_j} 
\end{equation}
By analogy with the ideal monoatomic gas model, and within the continuum limit, we define $E_T = \frac{E_{r}L}{a} D$, where $D$ is a distance in the phase space. 
The hypersphere containing all the micro-ensembles has a radius $D$ and dimension $2$, as the number of ripple is only given by the indexes $(n_j, m_j)$ (ie. one only needs these two indexes to describe a single $\mu$-state).	
The number $\Omega$ of $\mu$-states is therefore : 
\begin{equation}
\Omega = \frac{\mathcal{V}_{tot}}{\mathcal{V}_{\mu}} = \frac{\pi}{\Gamma(2)}\left (\frac{E_T}{E_{r}}\frac{L}{a}\right )^2 \frac{1}{\mathcal{V}_{\mu}}
\end{equation}
In phase space, the volume of a $\mu$-state, $\mathcal{V}_{\mu}$, is given by the distance between two consecutive neighbors $j$ and $j'$ : 
\begin{equation}
\mathcal{V}_{\mu} = \left	 (m_j-m_{j'} \right )\left	 (n_j-n_{j'} \right ) = 1
\end{equation}
This leads to :
\begin{equation}
\Omega = \frac{\pi}{\Gamma(2)}\left (\frac{E_T}{E_{r}}\frac{L}{a}\right )^2 
\end{equation}
It is therefore possible to define an entropy $\mathcal{S}$, introducing the constant $k$ :
\begin{equation}
\mathcal{S} = k \ln(\Omega) = 2k \ln (E_T) + k \ln  \left [ \frac{\pi E_{r}^2L^2}{\Gamma(2)a^2}\right ]
\end{equation}
Following Boltzmann theory, an analogue of micro-canonical temperature $\Theta$ is defined as : 
\begin{equation}
\frac{1}{\Theta} = \frac{\partial \mathcal{S}}{\partial E_T} = \frac{2k}{E_T}
\end{equation}
Therefore, for a given effective temperature $\Theta$, there is a fixed energy $E=2k\Theta$ for a ripple distribution.
Note : $k\Theta$ may be seen as the energy contribution for the fluctuations of the curvature of the grapheme flake. Also, $k\Theta$ can be seen as a ripple distribution in every direction.

\begin{figure}[htbp]
\begin{center}
		\includegraphics[width=\textwidth]{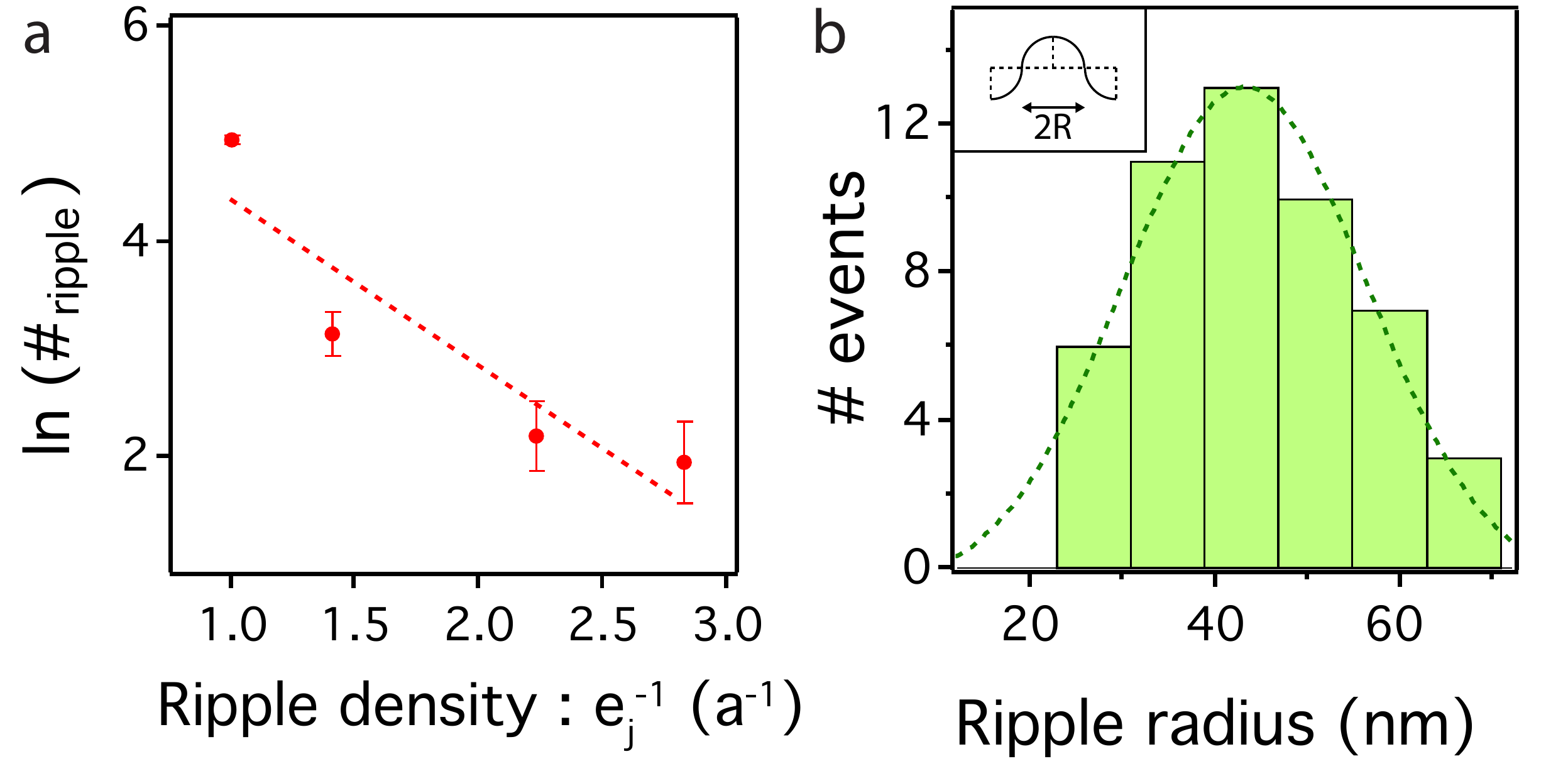}
	     \caption{\textbf{Statistical analysis of the distribution of ripples in graphene membrane. }\textbf{a}: Graphene ripples distribution as a function of the graphene ripples density $e_j^{-1}$ for different geometrical configurations (1$^{st}$, 2$^{nd}$, 3$^{rd}$ neighbor, etc). The experimental data has been extracted from one single nano-pillars square lattice (parameter a=1 $\mu$m). Red dashed line is a fit of data using \ref{eq:probapli}.\textbf{b}:  Graphene ripples diameter distribution. Data recorded from SEM micrographs. The central value is 42 nm (gaussian fit), and the distribution width is about 21 nm. Inset : sketch of a ripple cut. Graphene ripple is viewed as two half cylinders of opposite curvature.}\label{fig6}
\end{center}
\end{figure}

Therefore, it is possible to define the Bolztmann distribution :
\begin{equation}
P(E_T) = \Lambda \frac{e^{-\beta E_T}}{C} = \Lambda \frac{e^{-\beta \frac{E_rL}{e_j}}}{C}
\label{eq:probapli}
\end{equation}
where $\beta = (k \Theta)^{-1}$, $\Lambda$ is the degeneracy of the $j^{th}$ ripple density configuration, $C$ is the partition function normalizing the probability.
Statistical analysis of SEM micrographs of sample with the same pillars lattice parameter $a$ leads to the distribution of ripple lines linking 1$^{st}$, 2$^{nd}$, 3$^{rd}$, … neighbors  (cf. \ref{fig6} a).
This distribution reveals the probability $P(E_T)/\Lambda_j$ for each given ripple density $e_j^{-1}$.
Experimental results shown in \ref{fig6}a are in agreement with the numerical fit using \ref{eq:probapli} (dashed line); thus indicating that formation of graphene ripples onto periodic nano-pillars array is governed by pillars density as suggested by our model.

\section{Additional data}
\begin{figure}[htbp]
\begin{center}
		\includegraphics[width=9cm]{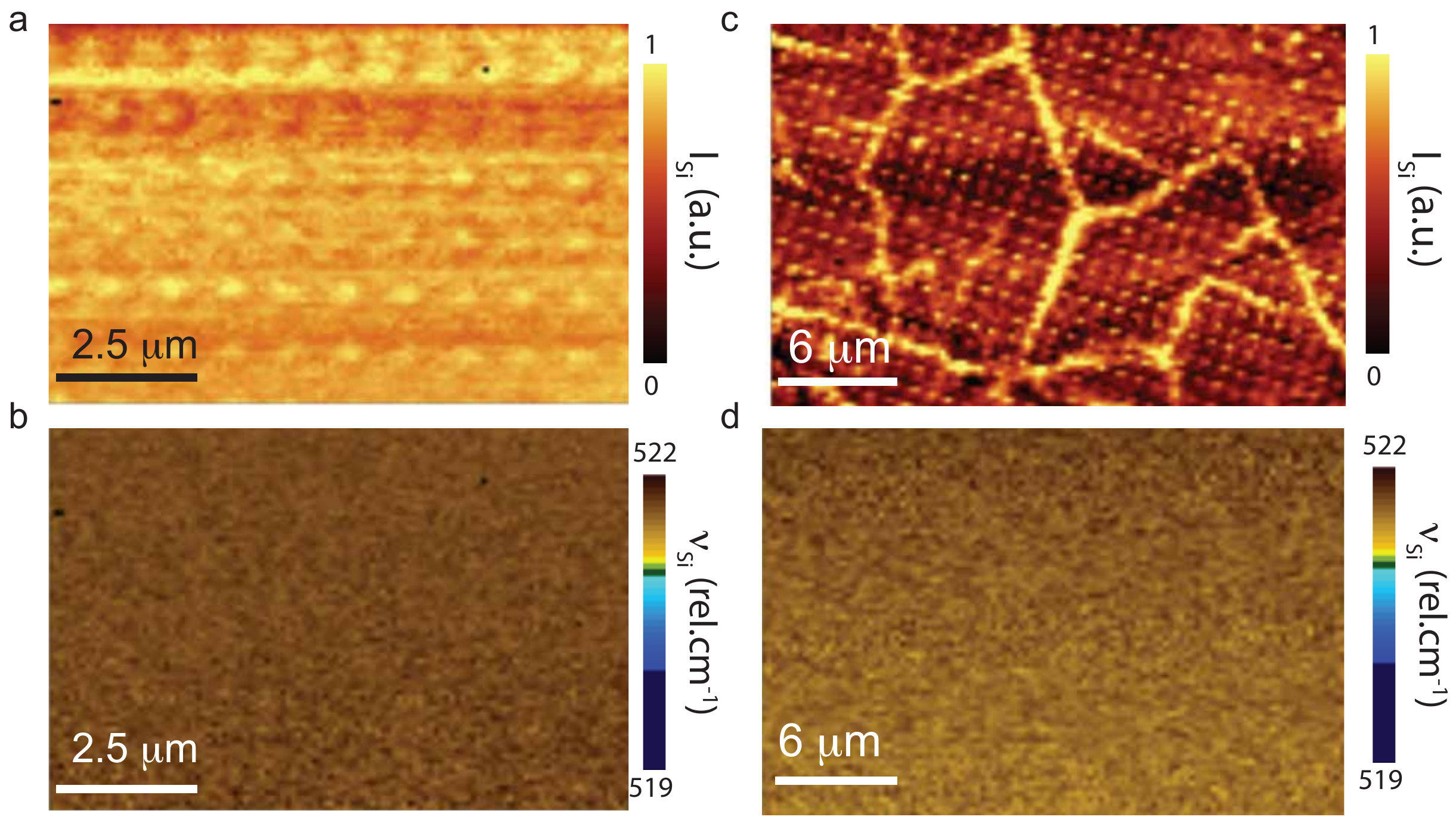}
	     \caption{\textbf{Raman response of the silicon mode before and after graphene transfer.} Raman map of the intensity (\textbf{a}) and of the frequency (\textbf{b}) of the Si-TO mode before the graphene transfer. After graphene transfer, the intensity (\textbf{c}) of the Si-TO mode is lower except at the graphene grain boundaries, which leaves the silicon surface exposed. The frequency (\textbf{d}) of the Si-TO mode is unchanged.}
\end{center}
\end{figure}

\begin{figure}[htbp]
\begin{center}
		\includegraphics[width=9cm]{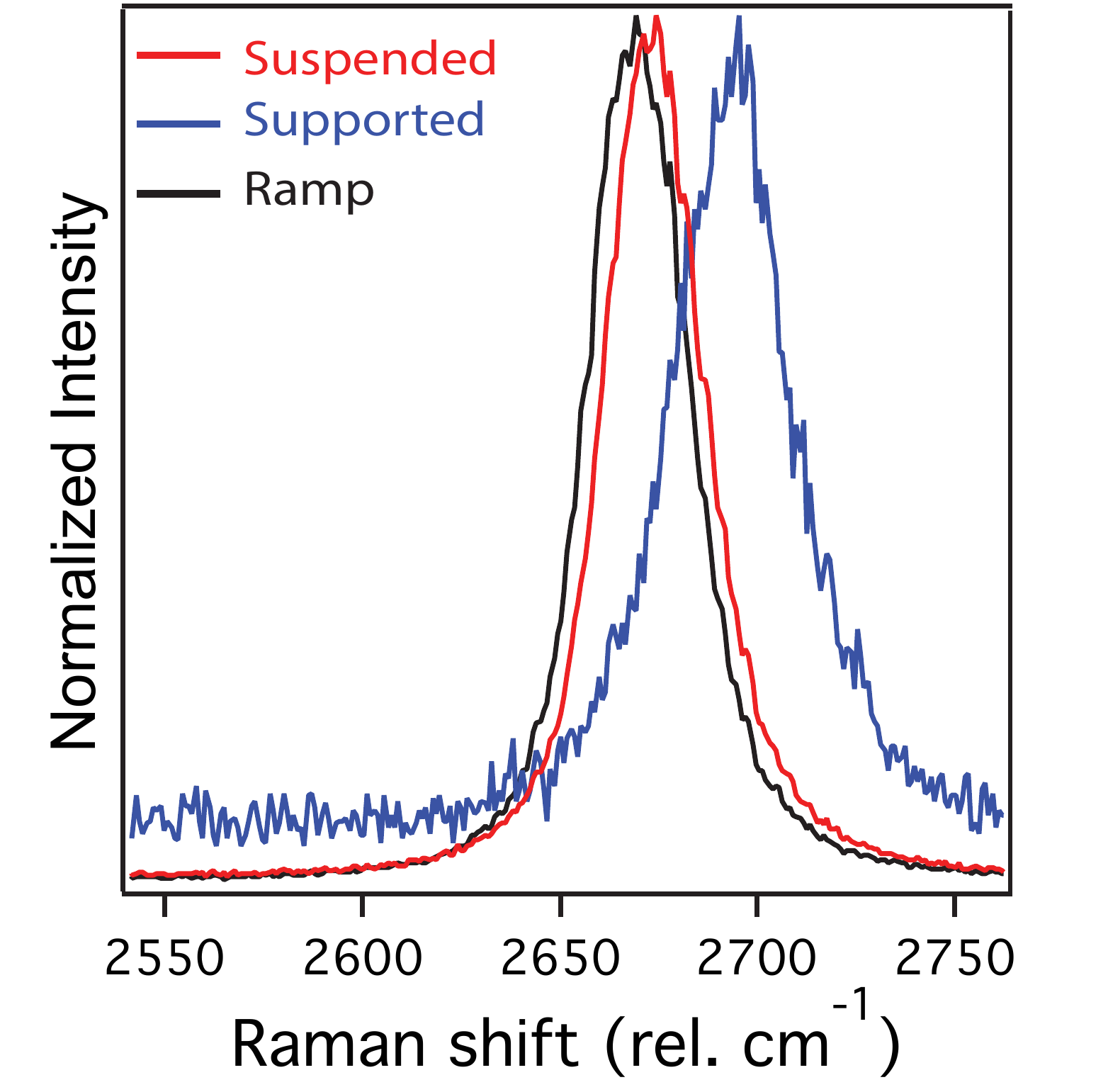}
	     \caption{\textbf{Raman 2D band for suspended and supported graphene.} Spectra of 2D band corresponding to the spots shown in \ref{fig5}-c. Spectra are fitted with two lorentzian functions, as suggested in reference \cite{Berciaud2013}, due to the coexistence of resonant inner and outer 2D processes. Fit results are presented in \ref{table}.}\label{figS7}
\end{center}
\end{figure}
\begin{table}
\begin{center}
\begin{tabular}{lccc}
\hline
	  & Suspended case (cm$^{-1}$) & Ramp (cm$^{-1}$) & Supported (cm$^{-1}$) \\
\hline
$\nu_{2D^-}$ & 2665.4 & 2661.5  & 2678.2 \\
$FWHM_{2D^-}$& 17.9 & 18.6 & 18.7 \\
$\nu_{2D^+}$ & 2676.2 & 2693.2 & 2673.3 \\
$FWHM_{2D^+}$ & 24.8& 23.6 & 31.2 \\

\hline
\end{tabular}
\end{center}
\caption{\textbf{2D band fit results.} 2D band profile fitted with two lorentzian functions, as suggested in reference\cite{Berciaud2013}. Raman spectra are shown in \ref{figS7}.}
\label{table}
\end{table}

\bibliography{BIB_Fakir_arxiv}
\end{document}